\documentclass[12pt,a4]{article}
\usepackage[dvips]{color,graphicx}
\usepackage{bm,psfrag,afterpage,color,amsmath, amssymb}

\makeatletter
\def\eqnarray{\stepcounter {equation}\let \@currentlabel =\theequation
\global \@eqnswtrue
\global \@eqcnt \z@ \tabskip \@centering \let \\=\@eqncr
$$\halign to \displaywidth \bgroup \@eqnsel \hskip \@centering
$\displaystyle \tabskip \z@ {##}$&\global \@eqcnt \@ne \hfil
${\mbox{}##\mbox{}}$\hfil &\global \@eqcnt \tw@
$\displaystyle \tabskip \z@ {##}$\hfil \tabskip \@centering
&\llap {##}\tabskip \z@ \cr}
\makeatother

\begin{document}

{\baselineskip = 8mm 

\begin{center}
\textbf{\LARGE Selection of tuning parameters in bridge regression models via Bayesian information criterion} \\[5mm]
\end{center}

\begin{center}
{\large Shuichi Kawano} \\[5mm]
\end{center}

\begin{center}
\begin{minipage}{14cm}
{
\begin{center}
{\it {\footnotesize 
Department of Mathematical Sciences, Graduate School of Engineering, \\ Osaka Prefecture University, 
1-1 Gakuen-cho, Sakai, Osaka 599-8531, Japan. \\
}}

\vspace{2mm}

skawano@ms.osakafu-u.ac.jp

\end{center}

\vspace{1mm} 

{\small {\bf Abstract:} \
We consider the bridge linear regression modeling, which can produce a sparse or non-sparse model. 
A crucial point in the model building process is the selection of adjusted parameters including a regularization parameter and a tuning parameter in  bridge regression models. 
The choice of the adjusted parameters can be viewed as a model selection and evaluation problem.
We propose a model selection criterion for evaluating bridge regression models in terms of Bayesian approach. 
This selection criterion enables us to select the adjusted parameters objectively. 
We investigate the effectiveness of our proposed modeling strategy through some numerical examples. 
}

\vspace{3mm}


\vspace{3mm}

{\small \noindent {\bf Mathematics Subject Classication:} 62J05, 62G05, 62F15.}
}

\end{minipage}
\end{center}

\baselineskip = 8mm

\vspace{3mm}


\section{Introduction}


With the appearance of high-throughput data of unexampled size and complexity,  statistical methods have increasingly become important. 
In particular, the linear regression models are widely used and fundamental tools in statistics. 
The parameters in the regression models are usually estimated by the ordinary least squares (OLS) or the maximum likelihood method. 
However, the models estimated by these methods often cause unstable estimators of the parameters and yield large prediction errors, when there exists in the multicollinearity in the regression models.

In order to overcome the problem, various penalized regression methods, e.g., the ridge regression (Hoerl and Kennard, 1970), the lasso (Tibshirani, 1996), the bridge regression (Frank and Friedman, 1993), the elastic net (Zou and Hastie, 2005), the SCAD (Fan and Li, 2001) and the MCP (Zhang, 2010), have been proposed. 
Among the penalized methods, we focus on the bridge linear regression method, which is the linear regression models estimated by the penalized method with the bridge penalty. 
An advantage of the bridge regression is to be able to produce a sparse model, which has received considerable attention in the high-dimensional data analysis that has exhaustibly studied in the late machine learning and statistical literature (see, e.g., B\"{u}hlmann and van de Geer, 2011), or a non-sparse model by controlling a tuning parameter included in the penalty function. 
Also, many researches (e.g., Armagan, 2009; Fu, 1998; Huang {\it et al.}, 2008; Knight and Fu, 2000) have showed that the bridge regression models are helpful from the practical and theoretical perspectives. 
Although the bridge regression is useful as seen above, there remains a problem of evaluating the bridge regression models, which leads to the selection of adjusted parameters involved in the constructed bridge regression models. 
For evaluating the models, the cross-validation (CV) is often utilized. 
The computational time of the CV, however, tends to be very large, and the high variability and tendency to undersmooth in CV are not negligible, since the selectors are repeatedly applied.

In this paper, we present a model selection criterion for evaluating the models estimated by the penalized maximum likelihood method with the bridge penalty from the viewpoint of Bayesian approach. 
The proposed criterion enables us to select appropriate values of the adjusted parameters in the bridge regression models objectively. 
Through some numerical studies, we investigate the performance of our proposed procedure.

This paper is organized as follows. 
Section 2 describes the bridge linear regression models with estimation algorithm. 
In Section 3, we introduce a model selection criterion derived from Bayesian viewpoint to choose some adjusted parameters in the models. 
Section 4 conducts Monte Carlo simulations and a real data analysis to examine the performance of our proposed strategy and to compare several types of criteria and methods. 
Some concluding remarks are given in Section 5.

\section{Bridge regression modeling}


\subsection{Preliminary}
\label{Preliminary}

Suppose that we have a data set $\{ (y_i, {\bm x}_i); i=1,\ldots,n \}$, where $y_i \in {\mathbb R}$ is a response variable and ${\bm x}_i = (x_{i1}, \ldots, x_{i p})^T$ denotes a $p$-dimensional covariate vector. 
Without loss of generality, it is assumed that the response is centered and the covariate is standardized, that is, 
\begin{eqnarray}
\sum_{i = 1}^n y_i = 0, \hspace{5mm} \sum_{i = 1}^n x_{ij} = 0, \hspace{5mm} \sum_{i = 1}^n x_{ij}^2 = n, \hspace{8mm} j=1,\ldots,p.
\end{eqnarray}
In order to capture a relationship between the response $y_i$ and the covariate vector ${\bm x}_i$, we consider the linear regression model 
\begin{eqnarray}
{\bm y} = X {\bm \beta} + {\bm \varepsilon},
\label{LinearModel}
\end{eqnarray}
where  ${\bm y} = (y_1, \ldots, y_n)^T$ is an $n$-dimensional response vector, $X$ is an $n \times p$ design matrix, ${\bm \beta} = (\beta_1, \ldots, \beta_p)^T$ is a $p$-dimensional coefficient vector and $\bm \varepsilon = (\varepsilon_1, \ldots, \varepsilon_n)^T$ is an $n$-dimensional error vector. 
In addition, we assume that the error $\varepsilon_i \ (i=1,\ldots,n)$ is independently distributed as the normal distribution with mean zero and variance $\sigma^2$.

From above some assumptions, we have a probability density function for the response ${\bm y}$ in the following:
\begin{eqnarray}
f(y_i | {\bm x}_i ; {\bm \theta}) = \frac{1}{\sqrt{2 \pi \sigma^2}} \exp \left[ - \frac{( y_i - {\bm x}_i^T {\bm \beta} )^2}{2 \sigma^2} \right], \quad i=1,\ldots,n,
\end{eqnarray}
where ${\bm \theta} = ({\bm \beta}^T, \sigma^2)^T$ is a parameter vector to be estimated. 
This leads to a log-likelihood function given by
\begin{eqnarray}
\ell ({\bm \theta}) = \sum_{i=1}^n \log f(y_i | {\bm x}_i ; {\bm \theta}) = - \frac{n}{2} \log (2 \pi \sigma^2) - \frac{1}{2 \sigma^2} \sum_{i=1}^n (y_i - {\bm x}_i {\bm \beta})^2.
\end{eqnarray}

\subsection{Estimation via the bridge penalty}
\label{Estimation}

The unknown parameter ${\bm \theta}$ is estimated by the penalized maximum likelihood method, that is, maximizing a penalized log-likelihood function
\begin{eqnarray}
\ell_{\lambda} ({\bm \theta}) = \ell ({\bm \theta}) - n \sum_{j=1}^p p_{\lambda} (\beta_j),
\label{Penalized}
\end{eqnarray}
where $p_{\lambda} (\cdot)$ is a penalty function and $\lambda \ (>0)$ is a regularization parameter. 
Until now, many penalty functions have been proposed: e.g., the $L_2$ penalty or the ridge penalty (Hoerl and Kennard, 1970) given by $p_{\lambda} (\beta) = \lambda \beta^2/2$, the $L_1$ penalty or the lasso penalty (Tibshirani, 1996) given by $p_{\lambda} (\beta) = \lambda |\beta|$, the elastic net penalty (Zou and Hastie, 2005) given by $p_{\lambda} (\beta) = \lambda \{ \alpha  \beta^2/2 + (1-\alpha) |\beta| \}$, where $\alpha \ (0\leq\alpha \leq 1)$ is a tuning parameter, the SCAD penalty (Fan and Li, 2001) given by $p^{\prime}_{\lambda} (\beta) = \lambda [ I (|\beta| \leq \lambda) + (a \lambda - |\beta|)_+ I(|\beta| > \lambda)/ \{ (a-1) \lambda\} ]$, where $a \ (> 2)$ is a tuning parameter and $(x)_+ = \max(0, x)$, and the MCP (Zhang, 2010) given by $p^{\prime}_{\lambda} (\beta) = (a \lambda - |\beta|)_+/a$, where $a \ (>0)$ is a tuning parameter. 
Note that the ridge penalty, the lasso penalty and the elastic net penalty are convex functions, while the SCAD penalty and the MCP are non-convex, and the lasso penalty, the elastic net penalty, the SCAD penalty and the MCP can produce sparse solutions for coefficient parameters, while the ridge penalty cannot. 
For more penalty functions, we refer to Antoniadis {\it et al.} (2011) and Lv and Fan (2009). 


In this paper, since we consider the bridge regression models, the formulation of the penalty $p_{\lambda} (\cdot)$ is $\lambda |\beta|^q/2$, called  the bridge penalty (Frank and Friedman, 1993), and then we obtain
\begin{eqnarray}
\ell_{\lambda, q} ({\bm \theta}) = \ell ({\bm \theta}) - \frac{n \lambda}{2} \sum_{j=1}^p |\beta_j|^q,
\label{Penalized2}
\end{eqnarray}
where $q \ (>0)$ is a tuning parameter. 
It is clear that the bridge penalty is the $L_1$ penalty when $q=1$ and is the $L_2$ penalty when $q=2$. 
Also, it is known that the bridge penalty yields sparse models if $0 < q \leq 1$, while the penalty yields non-sparse models if $q > 1$.

There are many researches about the bridge regression. 
Armagan (2009), Fu (1998) and Zou and Li (2008) proposed efficient algorithms for solving bridge regression models. 
Huang {\it et al.} (2008) and Knight and Fu (2000) showed asymptotic properties for linear regression models with the bridge penalty. 
Huang {\it et al.} (2009) and Park and Yoon (2011) extended the bridge penalty into the group bridge penalty, which is an extension of the group lasso penalty presented by Yuan and Lin (2006).

Since the bridge penalty is a convex function when $q \geq 1$, the Equation (\ref{Penalized2}) is a concave optimization problem. 
Hence, in order to obtain estimators of coefficient parameters, we can use usual optimization algorithms; e.g., the shooting algorithm (Fu, 1998).
However, since the bridge penalty is non-convex when $0 < q < 1$, the Equation (\ref{Penalized2}) is a non-concave optimization problem. 
Thus, we need to approximate the bridge penalty into a convex function. 
We apply the local quadratic approximation (LQA) introduced by Fan and Li (2001) for the bridge penalty.

For the LQA, under some conditions, the penalty function can be approximated at initial values ${\bm \beta}^{(0)} = (\beta_{1}^{(0)}, \ldots, \beta_{p}^{(0)})^T$ in the form
\begin{eqnarray}
|\beta_j|^q \approx |\beta_{j}^{(0)}|^q + \frac{q}{2} \frac{|\beta_{j}^{(0)}|^{q-1}}{|\beta_{j}^{(0)}|} (\beta_j^2 - \beta_{j}^{(0)2}), \qquad j=1,\ldots,p.
\end{eqnarray}
Then, the Equation (\ref{Penalized2}) can be expressed as
\begin{eqnarray}
\ell_{\lambda, q}^* ({\bm \theta}) = \ell ({\bm \theta}) - \frac{n \lambda q}{4} \sum_{j=1}^p |\beta_{j}^{(0)}|^{q-2} \beta_j^2.
\end{eqnarray}
This formulation is clearly a concave optimization problem since the bridge penalty is replaced with the quadratic function with respect to coefficient parameters $\beta_j \ (j=1,\ldots,p)$, and hence it is easy to obtain estimators of parameters $\bm \theta$. 
The estimators of parameters $\bm \theta$ can be derived according to the following algorithm:
\begin{description}
\item[Step1] Set the values of the regularization parameter $\lambda$ and the tuning parameter $q$, respectively. 
\item[Step2] Initialize ${\bm \beta}^{(0)} = (\beta_{1}^{(0)} \ldots, \beta_{p}^{(0)})^T$  and $\sigma^{(0)2}$. 
In our numerical studies, we set
\begin{eqnarray}
{\bm \beta}^{(0)} = (X^T X + n \gamma I_p)^{-1} X^T {\bm y}, \qquad \sigma^{(0)2} = 1,
\end{eqnarray}
where $\gamma = 10^{-5}$ and $I_p$ is a $p \times p$ identity matrix. 
\item[Step3] Update the coefficient vector $\bm \beta$ as follows:
\begin{eqnarray}
\hat{\bm \beta}^{(k+1)} = \{ X^T X + \Sigma_{\lambda, q} (\hat{\bm \beta}^{(k)}) \}^{-1} X^T {\bm y}, \qquad k=0,1,2,\ldots,
\end{eqnarray}
where $\Sigma_{\lambda, q} (\hat{\bm \beta}^{(k)}) = {\rm diag} (n \lambda \hat{\sigma}^{(k)2} q |\hat{\beta}_1^{(k)}|^{q-2}/4, \ldots, n \lambda \hat{\sigma}^{(k)2} q |\hat{\beta}_p^{(k)}|^{q-2}/4)$.
\item[Step4] Update the parameter $\sigma^2$ in the form
\begin{eqnarray}
\hat{\sigma}^{(k+1)2}= \frac{1}{n} ({\bm y} - X \hat{\bm \beta}^{(k+1)})^T ({\bm y} - X \hat{\bm \beta}^{(k+1)}).
\end{eqnarray}
\item[Step5] Repeat the Step3 into the Step4 until the following condition 
\begin{eqnarray}
| \hat{\bm \beta}^{(k+1)} - \hat{\bm \beta}^{(k)} | < \delta
\end{eqnarray}
is satisfied, where $\delta$ is an arbitrary small number (e.g., $10^{-5}$ in our numerical examples). 
\end{description}

From the procedures, we obtain the estimator $\hat{\bm \theta} = (\hat{\bm \beta}^T, \hat{\sigma}^2)^T$, and then it follows that we derive a statistical model
\begin{eqnarray}
f(y_i | {\bm x}_i ; \hat{\bm \theta}) = \frac{1}{\sqrt{2 \pi \hat{\sigma}^2}} \exp \left[ - \frac{( y_i - {\bm x}_i^T \hat{\bm \beta} )^2}{2 \hat{\sigma}^2} \right], \quad i=1,\ldots,n.
\end{eqnarray}
The statistical model includes some adjusted parameters, i.e., the regularization parameter $\lambda$ and the tuning parameter $q$. 
In order to choose these parameters objectively, we introduce a model selection criterion in terms of Bayesian approach.

\section{Model selection criteria}

\subsection{Proposed criterion}
Schwarz (1978) proposed the Bayesian information criterion (BIC) from the aspect of Bayesian theory. 
The BIC, however, is not applicable to models estimated by other methods except for the maximum likelihood method.
Konishi {\it et al.} (2004) extended the BIC such that it could be used for evaluating statistical models estimated by the penalized maximum likelihood method.


The Bayesian approach is to select the values of regularization parameter $\lambda$ and tuning parameter $q$ that maximizes the marginal likelihood. 
The marginal likelihood is calculated by integrating over the unknown parameter $\bm \theta$ and is defined by
\begin{eqnarray}
{\rm ML} =  \int \prod_{i=1}^{n} f ({y}_i | {\bm x}_i ; {\bm \theta}) \pi ({\bm \theta}) d {\bm \theta} = \int \prod_{i=1}^{n} f ({y}_i | {\bm x}_i ; {\bm \theta}) \pi ({\bm \beta} | \sigma^2) \pi ({\sigma^2}) d {\bm \theta},
\label{posterior-probability}
\end{eqnarray}
where $\pi ({\bm \theta}) = \pi ({\bm \beta} | \sigma^2) \pi ({ \sigma}^2)$ is the prior distribution of the parameter ${\bm \theta}$. 
In bridge regression models, the prior distribution $\pi ({\sigma^2})$ is assumed to be a non-informative prior distribution and  the prior distribution $\pi ({\bm \beta} | \sigma^2) = \pi ({\bm \beta})$  can be found in Fu (1998) as follows:
\begin{eqnarray}
\pi ({\bm \beta} | \lambda, q) &=& \prod_{j=1}^p \pi ({\beta}_j | \lambda, q) = \prod_{j=1}^p \frac{q 2^{- (1+1/q)} (n \lambda)^{1/q}}{\Gamma(1/q)} \exp \left\{ - \frac{n \lambda}{2} |\beta_j|^q  \right\},
\end{eqnarray}
where $\Gamma (\cdot)$ is the Gamma function. 

In general, it is difficult to evaluate the Equation (\ref{posterior-probability}), since we must often calculate a high-dimensional integral. 
Hence, some approximation methods are usually applied for the integral, for example, the Laplace approximation (Tierney and Kadane, 1986). 
However, in situations where some components of $\bm \beta$ are exactly zero with bridge approaches, the functional in the integral (\ref{posterior-probability}) is not differentiable at the origin, and then the approximation methods cannot be directly applied.

Let $\mathcal A = \{ j ; \hat{\beta}_j \neq 0 \}$ be active set of the parameter $\bm \beta$. 
In order to overcome the problem, we consider the partial marginal likelihood given by
\begin{eqnarray}
{\rm ML} \approx {\rm PML} =  \int \prod_{i=1}^{n} f ({y}_i | {\bm x}_i ; {\bm \theta}) \pi ({\bm \beta} | \lambda, q) d {\bm \theta}_{\mathcal A},
\label{posterior-probability2}
\end{eqnarray}
where ${\bm \theta}_{\mathcal A} = ({\bm \beta}_{\mathcal A}^T, \sigma^2)^T$. 
Here ${\bm \beta}_{\mathcal A} = (\beta_{k_1}, \ldots, \beta_{k_r})^T$, where we set ${\mathcal A} =\{ k_1, \ldots, k_r \}$ and $k_1 < \cdots < k_r$. 
The quantity is calculated by integrating over the unknown parameter ${\bm \theta}_{\mathcal A}$ included with the active set $\mathcal A$. 
Applying the Laplace approximation for the Equation (\ref{posterior-probability2}), we obtain
\begin{eqnarray}
 \int \prod_{i=1}^{n} f ({y}_{i} | {\bm x}_{i} ; {\bm \theta}) \pi ({\bm \beta} | \lambda, q) d {\bm \theta}_{\mathcal A}
= \frac{(2 \pi)^{|\mathcal A| + 1}}{n^{|\mathcal A| + 1} | { V} (\hat{\bm \theta}_{\mathcal A}) |^{1/2}} \exp \{ n v (\hat{\bm \theta}_\mathcal A) \} \left\{ 1 + O_p (n^{-1}) \right\}, 
\end{eqnarray}
where 
\begin{eqnarray}
v ({\bm \theta}) = \frac{1}{n} \log \left\{ \prod_{i=1}^{n} f ({y}_{i} | {\bm x}_{i} ; {\bm \theta}) \pi ({\bm \beta} | \lambda, q) \right\}, \quad { V} ({\bm \theta}) = - \frac{\partial^2 v ({\bm \theta})}{\partial {\bm \theta}_{\mathcal A} \partial {\bm \theta}^T_{\mathcal A}} \nonumber
\end{eqnarray}
and $\hat{\bm \theta}_{\mathcal A} = (\hat{\bm \beta}_{\mathcal A}^T, \hat{\sigma}^2)^T$, where $\hat{\bm \beta}_{\mathcal A}$ is the estimator of the coefficient ${\bm \beta}_{\mathcal A}$.

By taking the logarithm of the formula calculated by the Laplace approximation, Konishi {\it et al.} (2004) presented the generalized Bayesian information criteria (GBIC) to evaluate models estimated by the penalized maximum likelihood method. 
Uisng the result of Konishi {\it et al.} (2004, p. 30), we derive a model selection criterion
\begin{eqnarray}
{\rm GBIC} &=& n \log (2 \pi) + n \log \hat{\sigma}^2 + n - (|{\mathcal A}| + 1) \log \left( \frac{2 \pi}{n} \right) + \log |{ J}| \nonumber \\
& & {}  -2 |{\mathcal A}| \log q + 2 |{\mathcal A}| \left( 1 + \frac{1}{q} \right) \log 2- \frac{2 |{\mathcal A}|}{q} \log (n \lambda) + 2 |\mathcal A| \log \Gamma \left(\frac{1}{q} \right) + n \lambda \sum_{j \in {\mathcal A}} |\hat{\beta_j}|^q
, \nonumber \\
\label{GBIC}
\end{eqnarray}
where ${ J}$ is a $(|{\mathcal A}|+1) \times (|{\mathcal A}|+1)$ matrix given by
\begin{eqnarray}
{J} = {\displaystyle \frac{1}{n \hat{\sigma}^2}}\left(
\begin{array}{cc}
X_{\mathcal A}^T X_{\mathcal A} + n \lambda \hat{\sigma}^2 q (q-1) K & {\displaystyle \frac{1}{\hat{\sigma}^2}}X_{\mathcal A}^T \Lambda {\bm 1}_n\\
{\displaystyle \frac{1}{\hat{\sigma}^2}} {\bm 1}_n^T \Lambda X_{\mathcal A} & {\displaystyle \frac{n}{2 \hat{\sigma}^2}}\\
\end{array}
\right).
\end{eqnarray}
Here  ${\bm 1}_n = (1,\ldots,1)^T$ is an $n$-dimensional vector, $\Lambda = {\rm diag} (y_1 - {\bm x}_1^T \hat{\bm \beta}, \ldots, y_n - {\bm x}_n^T \hat{\bm \beta})$, $K = {\rm diag} (|\hat{\beta}_{k_1}|^{q-2}/2, \ldots, |\hat{\beta}_{k_r}|^{q-2}/2)$ and
\begin{eqnarray}
X_{{\mathcal A}} = [x_{ik}], \ \ i=1,\ldots,n; \ \ k \in {\mathcal A}.
\end{eqnarray}

We choose adjusted parameters, including values of regularization parameter $\lambda$ and tuning parameter $q$, from the minimizer of the GBIC in Equation (\ref{GBIC}).

\subsection{Other criteria}
\label{other criteria}
This section describes other selection criteria for choosing adjusting parameters included in bridge regression models.

\subsubsection{Modified AIC and modified BIC}
As an approximation of the effective degrees of freedom in the model selection theory, Hastie and Tibshirani (1990) proposed to use the trace of the hat matrix. 
In bridge regression models, the hat matrix is given by $S=X_{\mathcal A} (X^T_{\mathcal A} X_{\mathcal A}  + \Sigma_{\lambda, q} (\hat{\bm \beta}_{\mathcal A}))^{-1} X_{\mathcal A}^T$, where $\Sigma_{\lambda, q} (\hat{\bm \beta}_{\mathcal A}) = {\rm diag} (n \lambda \hat{\sigma}^{2} q |\hat{\beta}_{k_1} |^{q-2}/4, \ldots, n \lambda \hat{\sigma}^{2} q |\hat{\beta}_{k_r} |^{q-2}/4)$.
By replacing the number of parameters in AIC (Akaike, 1974) and BIC (Schwarz, 1978) with the trace of the hat matrix $S$, we obtain the modified AIC and modified BIC, respectively, 
\begin{eqnarray}
{\rm mAIC} &=& -2 \sum_{i=1}^n \log f (y_i | {\bm x}_i ; \hat{\bm \theta}) + 2 {\rm tr} S, \\
{\rm mBIC} &=& -2 \sum_{i=1}^n \log f (y_i | {\bm x}_i ; \hat{\bm \theta}) +  ({\rm tr} S) \log n.
\end{eqnarray}
A problem may arise in theoretical justification for the use of the bias-correction term, since the AIC and the BIC only cover statistical models estimated by the maximum likelihood method.

\subsubsection{Bias corrected AIC}
Hurvich and Tsai (1989) and Sugiura (1978) proposed an improved version of AIC in the context of linear regression models and autoregressive time series models estimated by the maximum likelihood method. 
Hurvich {\it et al.} (1998) presented to replace the number of parameters in the improved version of AIC with the trace of the hat matrix and introduced the criterion
\begin{eqnarray}
{\rm AICc} = -2 \sum_{i=1}^n \log f (y_i | {\bm x}_i ; \hat{\bm \theta}) + \frac{2 n ({\rm tr} S + 1)}{n - {\rm tr }S -2}.
\end{eqnarray}

\subsubsection{Cross-validation and generalized cross-validation}
The cross-validation evaluates a statistical model for each observation by using the remaining data . 
Let $\hat{y}^{(-i)}$ be a regression response value estimated by the observed data except $(y_i, {\bm x}_i)$. 
The cross-validation criterion is then
\begin{eqnarray}
{\rm CV} = \frac{1}{n} \sum_{i=1}^n \left( y_i - \hat{y}^{(-i)} \right)^2 = \frac{1}{n} \sum_{i=1}^n \left( \frac{y_i - \hat{y}_i}{1 - s_{ii}} \right)^2,
\label{CV}
\end{eqnarray}
where $\hat{y}_1, \ldots, \hat{y}_n$ are fitted values and $s_{ii}$ is an $i$-th diagnal element of the hat matrix $S$.

Craven and Wahba (1979) proposed the generalized cross-validation by replacing the value $s_{ii}$ in the Equation (\ref{CV}) with the trace of the hat matrix as follows:
\begin{eqnarray}
{\rm GCV}  = \frac{1}{n} \sum_{i=1}^n \left( \frac{y_i - \hat{y}_i}{1 - {\rm tr} S/n} \right)^2.
\label{GCV}
\end{eqnarray}

\subsubsection{Extended information criterion}
Let $\{ (y_i^{(b)}, {\bm x}_i^{(b)}) ; i=1,\ldots,n\} \ (b=1,\ldots,B)$ be the $b$-th bootstrap sample by resampling, and $\hat{\bm \theta}^{(b)}$ be the bridge estimator based on the  $b$-th bootstrap sample. 
The extended information criterion proposed by Ishiguro {\it et al.} (1997) is then defined by
\begin{eqnarray}
{\rm EIC} = -2 \sum_{i=1}^n \log f (y_i | {\bm x}_i ; \hat{\bm \theta}) + \frac{2}{B} \sum_{b=1}^B \left\{ \log f (y_i^{(b)} | {\bm x}_i^{(b)} ; \hat{\bm \theta}^{(b)}) - \log f (y_i | {\bm x}_i ; \hat{\bm \theta}^{(b)}) \right\}.
\end{eqnarray}
In our numerical experiments, $B$ is set to 100.

\section{Numerical results}

In order to show the efficiency of our proposed modeling strategy, we conducted some numerical examples. 
Monte Carlo simulations and analysis of real data are given to illustrate the proposed bridge modeling procedure. 

\subsection{Simulated examples}

We performed a simulation study to validate our proposed modeling procedure. 
The simulation has five settings, and the design matrix $X$ was generated from a multivariate normal distribution with mean zero and variance 1 for Settings 1, 2, 3 and 4, and then the correlation structure was given at each setting. 
The response vector $\bm y$ is generated from the true regression model
\begin{eqnarray}
{\bm y} = X {\bm \beta} + {\bm \varepsilon}, \qquad {\bm \varepsilon} \sim N ({\bm 0}, \sigma^2 I_n),
\end{eqnarray}
where $I_n$ is an $n \times n$ identity matrix. 
Our five simulation settings are given as follows:
\begin{itemize}
\item Setting 1 : The training data and the test data consisted of 20 observations and 200 observations, respectively. 
The true parameter was $\bm \beta = (3,15,7.5,5,2,0,0,0,0,0)^T$ and $\sigma=3$. 
The pairwise correlation between ${\bm x}_i$ and ${\bm x}_j$ was ${\rm cor} ({\bm x}_i, {\bm x}_j) =0.5^{|i-j|}$. 
This setting is the sparse case. 
\item Setting 2 : This setting is the same as the Setting 1 except for $\beta_j = 10 \ (j=1,\ldots, 10)$. 
That is, the Setting 2 is the dense case. 
\item Setting 3 : This setting is also the same as the Setting 1. 
However, the true parameter was ${\bm \beta} = (5,0,0,0,0,0,0,0)^T$ and $\sigma=2$. 
In this model, we consider the sparse case. 
\item Setting 4 : 100 observations and 400 observations were generated for the training data and the test data, respectively. 
We set
\begin{eqnarray}
{\bm \beta} = (\underbrace{0, \ldots,0}_{10}, \underbrace{5, \ldots,5}_{10}, \underbrace{0, \ldots,0}_{10}, \underbrace{3, \ldots,3}_{10})^T
\end{eqnarray}
and $\sigma=3$. 
The pairwise correlation between ${\bm x}_i$ and ${\bm x}_j$ was ${\rm cor} ({\bm x}_i, {\bm x}_j) =0.95^{|i-j|}$. 
This model also consider the sparse pattern. 
\item Setting 5 : The generating procedure of the training data and test data is the same as the Setting 4. 
The true parameter was
\begin{eqnarray}
{\bm \beta} = (\underbrace{10, \ldots,10}_{35}, \underbrace{0, \ldots,0}_{5})^T
\end{eqnarray}
and $\sigma=3$. 
The design matrix $X$ was generated as follows:
\begin{eqnarray}
&& x_{ij} = Z_k + \varepsilon_j, \quad Z_k \sim N(0,1), \quad j=5k-4, \ldots,5k, \ k=1,\ldots,7 \ {\rm for \ all} \ i, \\
&& x_{ij} \sim N(0,1), \quad j=36, \ldots, 40 \ {\rm for \ all} \ i,
\end{eqnarray}
where $\varepsilon_j$ were identically distributed as $N(0, 0.01)$ for $j=1,\ldots,35$. 
This model was also the sparse case. 
\end{itemize}

We fitted the bridge regression models to the simulated data. 
The regularization parameter $\lambda$ and the tuning parameter $q$ in the bridge penalty were selected by GBIC, mAIC, mBIC, AICc, CV, GCV and EIC, where we set the candidate values of $\lambda$ and $q$ to $\{ 10^{-0.1 i + 3} ; i=1,\ldots,100\}$ and $\{ 0.1, 0.4, 0.7, 1, 1.3, 1.7, 2, 2.3, 2.7 \}$, respectively.

We computed the mean squared error (MSE) defined by ${\rm MSE} = \sum_{i=1}^n (\hat{y}_i - y_i^*)^2/n$, where $y_1^*, \ldots, y_n^*$ denote test data for the response variable generated from the true model. 
Also, the means and standard deviations of the adjusted parameters $\lambda$ and $q$ were computed. 
The simulation results were obtained by averaging over 100 Monte Carlo trials, which are shown in Tables {\ref{simulation1}}, {\ref{simulation2}}, {\ref{simulation3}}, {\ref{simulation4}} and {\ref{simulation5}}. 
The values in parentheses indicate standard deviations for the means.

\begin{table}[htbp]
\begin{center}
\caption{Comparisons of the mean squared error (MSE) based on various criteria for the Setting 1. 
Figures in parentheses give estimated standard deviations.}
\vspace{5mm}
\begin{tabular}{lcccccccc}
\hline
          & GBIC  & mAIC & mBIC & AICc & CV & GCV & EIC \\ \hline
MSE & 15.67   & 17.13 & 16.18 &  15.80 &  16.27 &  16.02 &  17.89  \\
 & (6.12)   & (6.83) & (6.15) &  (5.99) &  (6.08) &  (6.00) &  (11.39)  \\
$\log_{10} (\lambda)$ & --1.137   & --0.803 & --0.544 &  --0.412 &  --0.599  &  --0.593 &  --0.625  \\
 & (0.4138)   & (0.3343) & (0.2198) &  (0.1849) &  (0.3875) &  (0.2425) &  (0.7058)  \\
$q$ & 0.598   & 0.946 & 0.805 &  0.745 &  0.832 &  0.841 &  0.890  \\
 & (0.2670)   & (0.3229) & (0.2500) &  (0.1777) &  (0.3675) &  (0.2778) &  (0.2455)  \\
\hline
\end{tabular}
\label{simulation1}
\end{center}
\end{table}
\begin{table}[htbp]
\begin{center}
\caption{Comparisons of the mean squared error (MSE) based on various criteria for the Setting 2. 
Figures in parentheses give estimated standard deviations.}
\vspace{5mm}
\begin{tabular}{lcccccccc}
\hline
 & GBIC  & mAIC & mBIC & AICc & CV & GCV & EIC \\ \hline
MSE & 20.72   & 21.23 & 21.42 &  24.02 &  21.82 &  21.63 &  81.06  \\
 & (8.03)   & (8.43) & (8.67) &  (11.95) &  (10.16) &  (8.90) &  (240.06)  \\
$\log_{10} (\lambda)$ & --4.148   & --1.106 & --1.007 &  --0.932 &  --0.983 &  --0.963 &  --2.518  \\
 & (0.084)   & (0.975) & (0.972) &  (0.894) &  (1.557) &  (0.942) &  (0.369)  \\
$q$ & 2.700   & 1.140 & 1.176 &  1.372 &  1.289 &  1.201 &  2.560  \\
 & (0.000)   & (0.639) & (0.638) &  (0.652) &  (1.013) &  (0.639) &  (0.243)  \\
 \hline
\end{tabular}
\label{simulation2}
\end{center}
\end{table}
\begin{table}[htbp]
\begin{center}
\caption{Comparisons of the mean squared error (MSE) based on various criteria for the Setting 3. 
Figures in parentheses give estimated standard deviations.}
\vspace{5mm}
\begin{tabular}{lcccccccc}
\hline
 & GBIC  & mAIC & mBIC & AICc & CV & GCV & EIC \\ \hline
MSE & 4.986   & 5.836 & 5.213 &  5.068 &  5.470 &  5.551 &  5.500  \\
 & (1.158)   & (1.744) & (1.456) &  (1.385) &  (1.488) &  (1.549) &  (1.753)  \\
$\log_{10} (\lambda) $ & --0.741   & --0.550 & --0.158 &  --0.352 &  --0.498 &  --0.477 &  -0.618  \\
 & (0.3900)   & (0.3672) & (0.4311) &  (0.2921) &  (0.7602) &  (0.2714) &  (0.9798)  \\
$q$ & 0.466 & 0.844 & 0.556 &  0.565 &  0.652 &  0.754 &  0.778  \\
 & (0.2507)   & (0.4699) & (0.4522) &  (0.3447) &  (0.5003) &  (0.4377) &  (0.4760)  \\
\hline
\end{tabular}
\label{simulation3}
\end{center}
\end{table}
\begin{table}[htbp]
\begin{center}
\caption{Comparisons of the mean squared error (MSE) based on various criteria for the Setting 4. 
Figures in parentheses give estimated standard deviations.}
\vspace{5mm}
\begin{tabular}{lcccccccc}
\hline
 & GBIC  & mAIC & mBIC & AICc & CV & GCV & EIC \\ \hline
MSE & 11.76   & 11.93 & 12.21 &  11.92 &  11.87 &  11.87 &  11.54  \\
 & (1.199)   & (1.231) & (1.375) &  (1.254) &  (1.247) &  (1.247) &  (0.994)  \\
$\log_{10} (\lambda)$ & --2.094   & --0.788 & --0.548 &  --0.664 &  --0.710 &  --0.699 &  --1.834  \\
 & (0.2173)   & (0.2818) & (0.0559) &  (0.0785) &  (0.1267) &  (0.1010) &  (0.3590)  \\
$q$ & 0.874   & 1.030 & 1.000 &  1.003 &  1.009 &  1.006 &  1.890  \\
 & (0.1715)   & (0.0904) & (0.0000) &  (0.0300) &  (0.0514) &  (0.0422) &  (0.3729)  \\
 \hline
\end{tabular}
\label{simulation4}
\end{center}
\end{table}
\begin{table}[htbp]
\begin{center}
\caption{Comparisons of the mean squared error (MSE) based on various criteria for the Setting 5. 
Figures in parentheses give estimated standard deviations.}
\vspace{5mm}
\begin{tabular}{lcccccccc}
\hline
 & GBIC  & mAIC & mBIC & AICc & CV & GCV & EIC \\ \hline
MSE & 14.39   & 14.62 & 15.41 &  14.91 &  14.72 &  14.73 &  10.95  \\
 & (1.566)   & (1.861) & (2.028) &  (1.890) &  (1.868) &  (1.872) &  (1.001)  \\
$\log_{10} (\lambda)$  & --3.768   & --0.945 & --0.779 &  --0.858 &  --0.910 &  --0.901 &  --2.925  \\
 & (1.117)   & (0.107) & (0.068) &  (0.066) &  (0.092) &  (0.088) &  (0.527)  \\
$q$ & 1.827   & 1.009 & 1.000 &  1.000 &  1.006 &  1.006 &  2.34  \\
 & (0.8610)   & (0.0514) & (0.0000) &  (0.0000) &  (0.0422) &  (0.0422) &  (0.3613)  \\
\hline
\end{tabular}
\label{simulation5}
\end{center}
\end{table}

The simulation results are summarized as follows. 
In the Settings 1, 2 and 3, all criteria provide an appropriate value of the tuning parameter $q$: i.e., the tuning parameter $q $ is larger than 1 when the structure of the coefficient parameter ${\bm \beta}$ is dense and $0 < q \leq 1$ is given when the structure of the coefficient parameter ${\bm \beta}$ is sparse. 
For the Setting 4, the GBIC and the mBIC yield sparse solutions for coefficient vectors $\bm \beta$, while other criteria produce dense solutions. 
In the fifth setting, the mBIC and the AICc can select the appropriate value of the tuning parameter $q$, whereas other criteria including the GBIC, which is the proposed criterion in this paper,  do not. 
However, the GBIC is superior to other criteria in almost all cases in the sense of minimizing the MSE and MSEs for the GBIC have smaller standard deviations among various criteria except for the EIC in the Settings 4 and 5. 
Note that the EIC appears to be unstable, since the criterion provides the worst MSEs in the Settings 1 and 2 while the smallest MSEs are certainly given in the Settings 4 and 5. 
In addition, the EIC requires much computational load, and hence our proposed criterion GBIC seems to be useful from the viewpoints of minimization of the MSEs and computational times.

\begin{figure}[htbp]
\centering
\includegraphics[width=6.3cm,height=6.3cm]{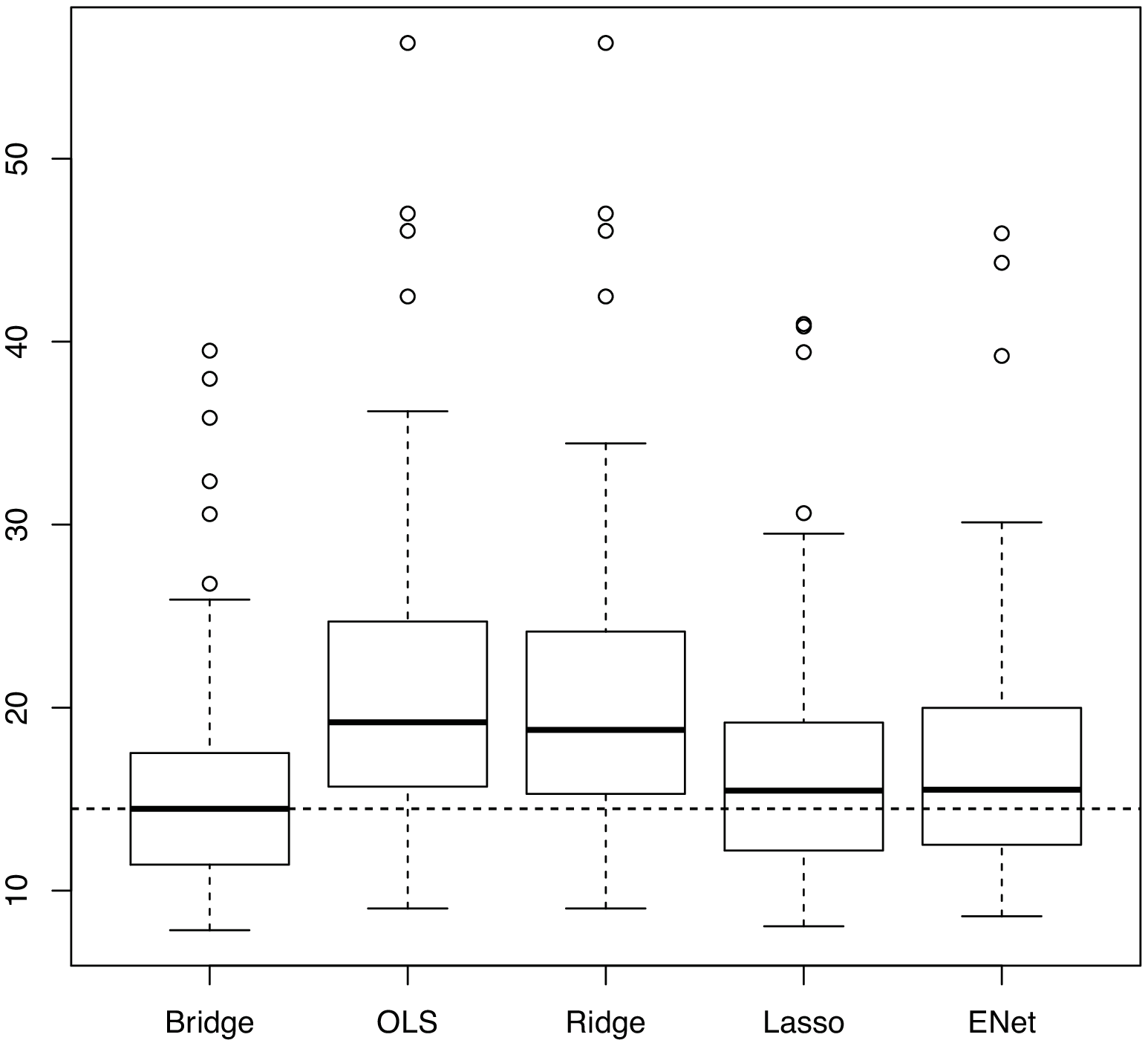}
\hspace{0.5cm} 
\includegraphics[width=6.3cm,height=6.3cm]{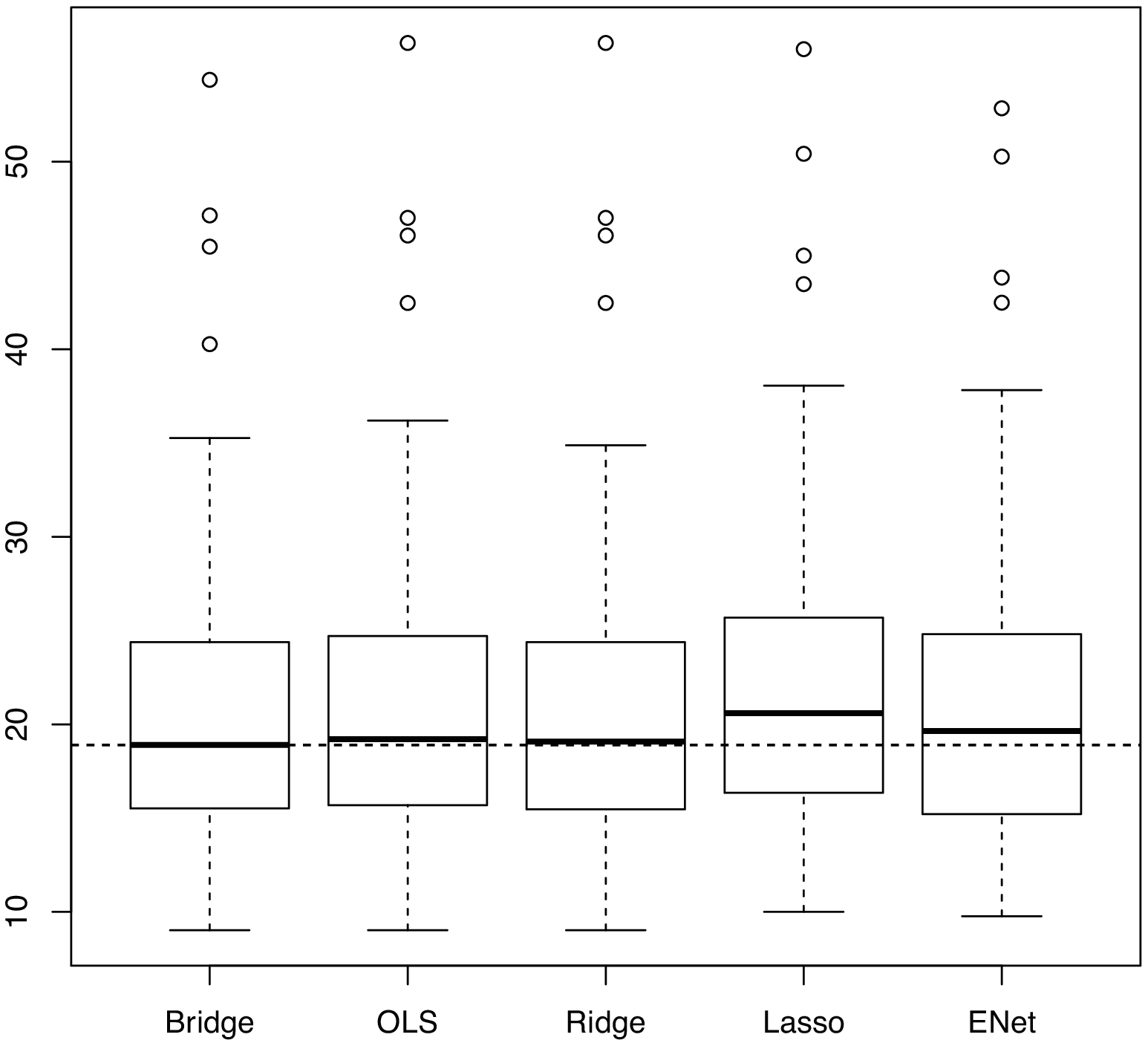}\\[0.5cm] 
\includegraphics[width=6.3cm,height=6.3cm]{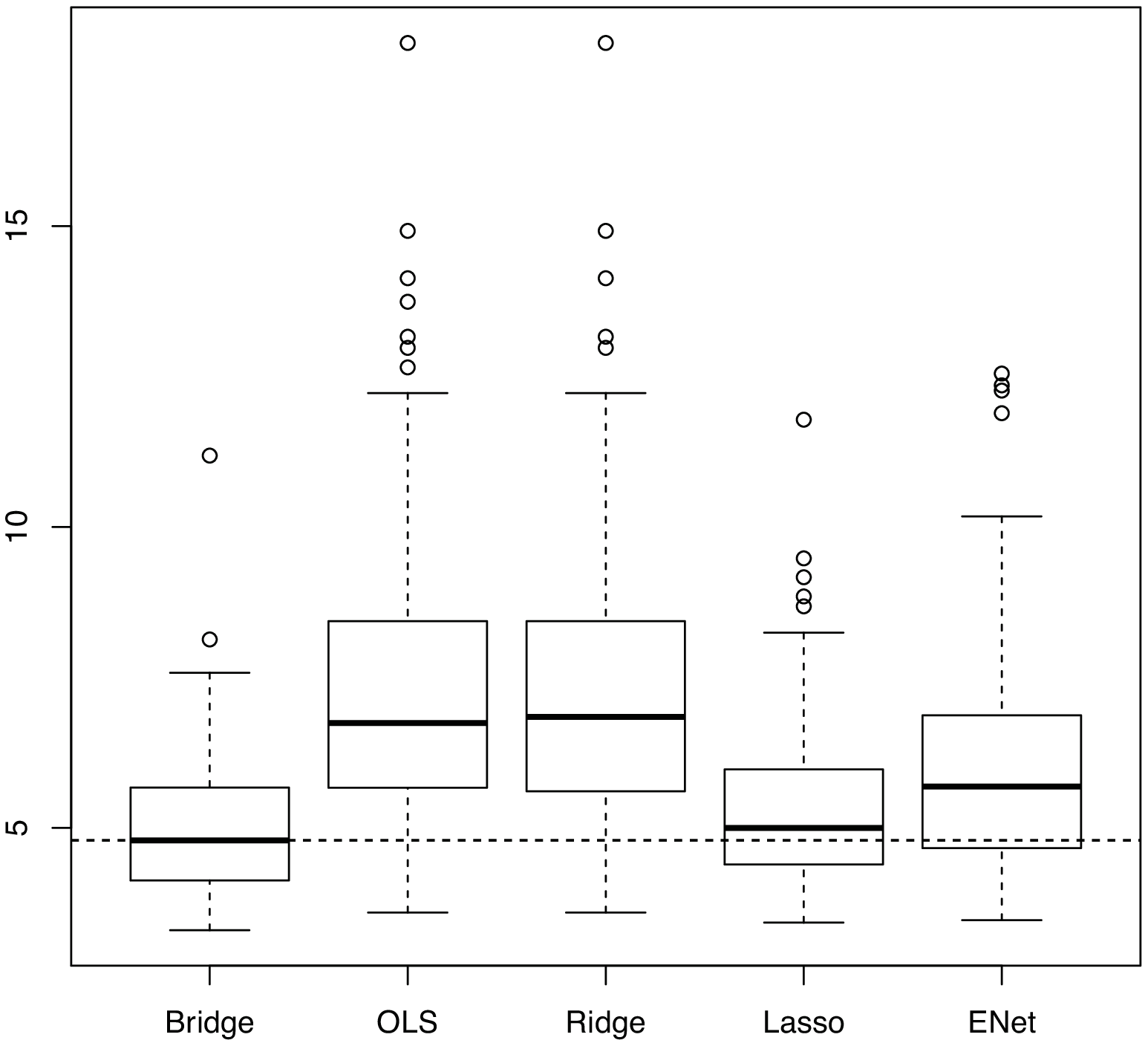}
\hspace{0.5cm} 
\includegraphics[width=6.3cm,height=6.3cm]{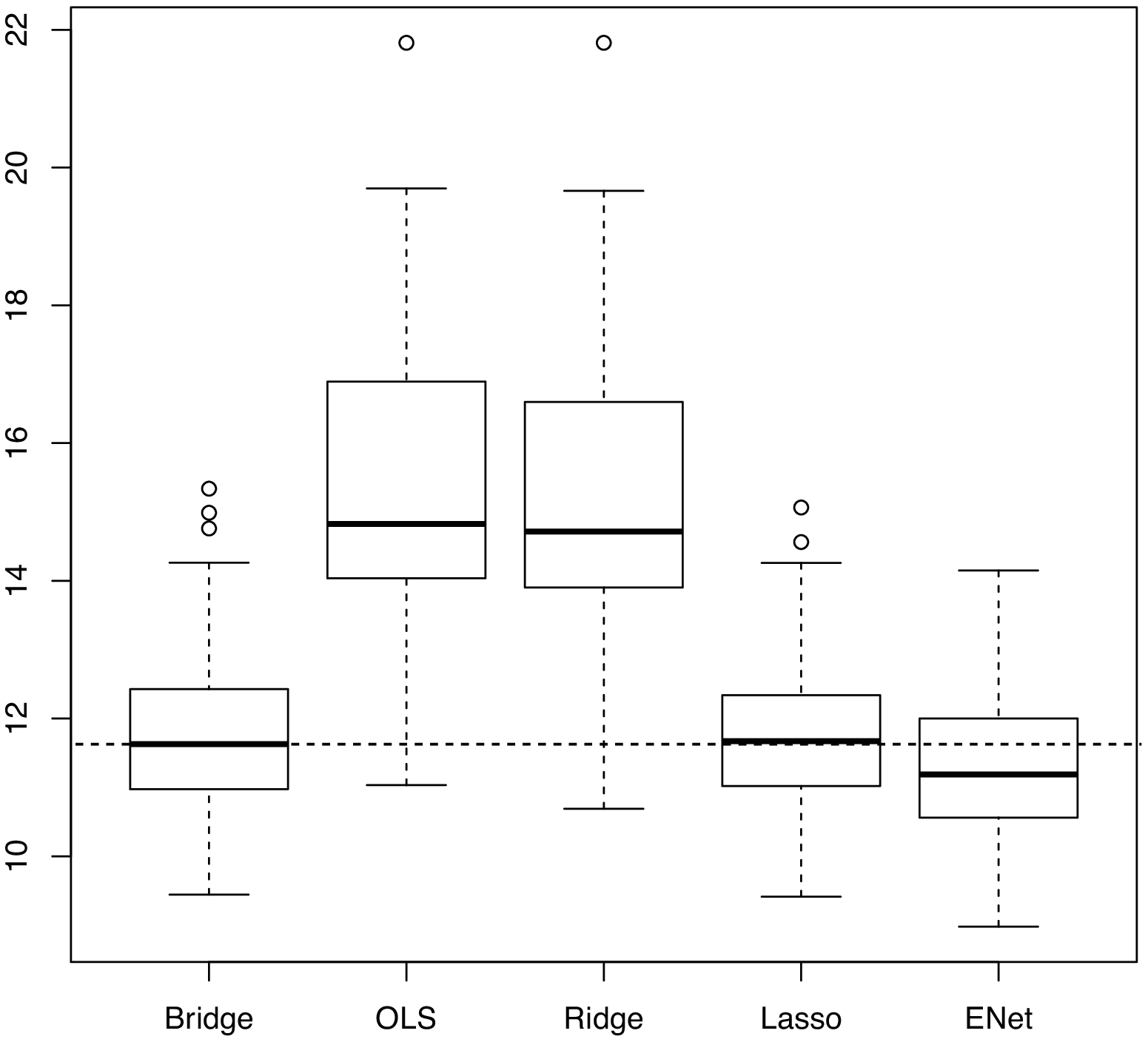}\\[0.5cm] 
\includegraphics[width=6.3cm,height=6.3cm]{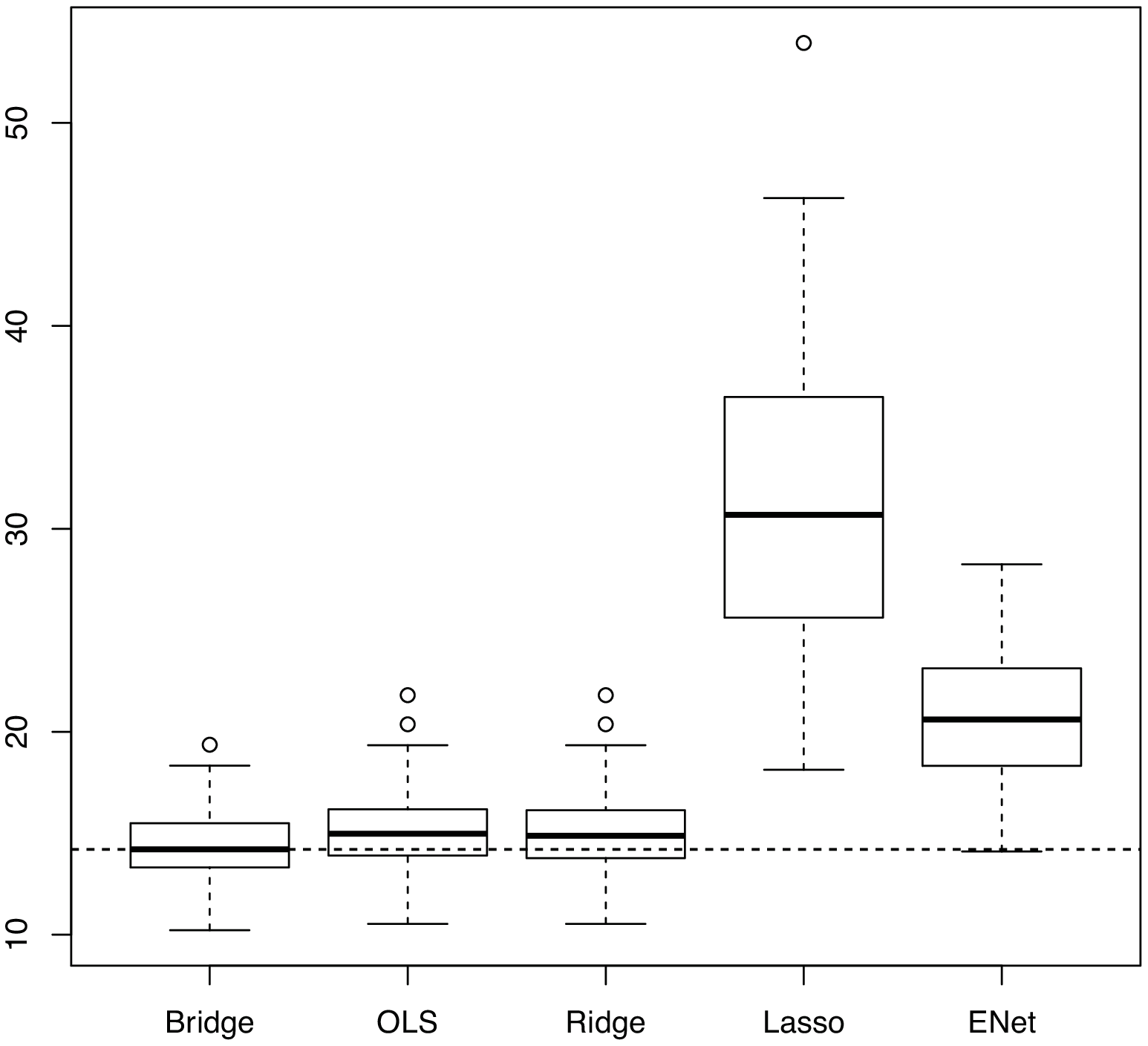}
\caption{Boxplots of the MSEs. 
The left top panel shows the result for the Setting 1, the right top panel that for the Setting 2, the left bottom panel that for the Setting 3, the right bottom panel that for the Setting 4 and the center bottom panel that for the Setting 5. }
\label{boxplot1}
\end{figure}

We also compared the bridge regression models with the GBIC to OLS, ridge, lasso and elastic net (ENet). 
An adjusted parameter included in ridge regression was selected by the leave-one-out cross-validation, and adjusted parameters involved in lasso and ENet were selected by the five-fold cross-validation. 
In order to evaluate the performance of each model, we computed the MSE, and described boxplots  of the values for the 100 trials of Monte Carlo simulations. 
Figure \ref{boxplot1} shows the boxplots of the MSEs. 
In almost situations, our proposed bridge regression modeling may perform well; i.e., it produces a relatively small median with small variance. 

\subsection{Analysis of real data}

We applied the bridge regression modeling evaluated by the GBIC to the pollution data set. 
This data were analyzed by McDonald and Schwing (1973), Liu {\it et al.} (2006) and Park and Yoon (2011). 
The data set consists of 60 observations and 15 covariates. 
The response variable is the total age-adjusted mortality rate obtained for the years 1059--1961 for 201 Standard Metropolitan Statistical Area. 
The data set is available from the {\ttfamily SMPracticals} package in the software {\ttfamily R}.

In order to validate prediction errors, we randomly divide the data set into 40 training data and 20 test data. 
Using the training data set, we constructed the regression models with the bridge penalty. 
The values of the regularization parameter $\lambda$ and the tuning parameter $q$ were chosen by using the GBIC. 
Here, we set the candidate values of $\lambda$ and $q$ into $\{ 10^{-0.1 i + 3} ; i=1,\ldots,100\}$ and $\{ 0.1, 0.25, 0.4, 0.55, 0.7, 0.85, 1, 1.3, 1.7, 2 \}$, respectively. 
The selected values of adjusted parameters were $\lambda=0.007943$ and $q=0.7$.

\begin{table}[t]
\begin{center}
\caption{Prediction errors for pollution data set.}
\vspace{5mm}
\begin{tabular}{lcccccc}
\hline
Method & bridge  & OLS & ridge & lasso & ENet \\ \hline
Prediction error & 1663.516   & 1822.312 & 1817.689 &  1735.713 &  1720.655  \\
 \hline
\end{tabular}
\label{pollution1}
\end{center}
\end{table}

\begin{table}[t]
\begin{center}
\caption{Selected variables for pollution data set.}
\vspace{5mm}
\begin{tabular}{ll}
\hline
Method & Selected variables   \\ \hline
McDonald and Schwing & (1,2,6,8,9,14)   \\
Luo {\it et al.} & (1,2,6,9,14)   \\
Park and Yoon (LQA) & (1,2,3,6,8,9,14)   \\
Park and Yoon (LLA) & (1,2,3,6,7,8,9,14,15)   \\
lasso & (1,2,6,8,9,14) \\
ENet & (1,2,6,8,9,14) \\
bridge with GBIC &(1,8,9,14) \\
 \hline
\end{tabular}
\label{pollution2}
\end{center}
\end{table}

We compared the performance of our modeling procedure with that of OLS, ridge, lasso and Enet. 
Table \ref{pollution1} summarizes the prediction errors by these methods. 
We observe that the bridge regression model outperforms other methods. 
Table \ref{pollution2} is the selected variables using the entire data in the pollution data set. 
Lasso and ENet choose the same variables with McDonald and Schwing (1973), and our proposed method has the smallest model among them. 
From these descriptions, we conclude that at least variables 1, 8, 9 and 14 may be relevant with the response variable, since the variables are included in all methods. 

\section{Concluding remarks}
\label{concluding remarks}

In this paper, we have considerd the problem of evaluating linear regression models estimated by the penalized maximum likelihood method with the bridge penalty. 
In order to select the optimal values of the adjusted parameters including the regularization parameter in the penalized maximum likelihood function and the tuning parameter in the bridge penalty, we have proposed a model selection criterion in terms of Bayesian theory. 
Monte Carlo simulations and analyzing a real data have showed that our proposed modeling procedure performs well in various situations from the  viewpoint of yielding relatively lower prediction errors than previously developed criteria and methods. 
The future work is to apply the proposed procedure into high-dimensional data sets and extend our models in the framework of generalized linear models.



{}


\begin{thebibliography}{}


\bibitem{} {Akaike, H. (1974). A new look at the statistical model identification. \textit{IEEE Transactions on Automatic Control}, \textbf{AC-19}, 716--723.}


\bibitem{} {Antoniadis, A., Gijbels, I. and Nikolova, M. (2011). Penalized likelihood regression for generalized linear models with non-quadratic penalties. {\it Annals of the Institute of Statistical Mathematics}, {\bf 63}, 585--615. }


\bibitem{} {Armagan, A. (2009). Variational bridge regression. {\it Proceedings of the 12th International Conference on Artificial Intelligence and Statistics}, {\bf 5}, 17--24. }


\bibitem{} {B\"{u}hlmann, P. and van de Geer, S. (2011). {\it Statistics for High-Dimensional Data}.  New York: Springer.}


\bibitem{} {Craven, P. and Wahba, G. (1979). Smoothing noisy data with spline functions. {\it Numerische Mathematik}, {\bf 31}, 377--403.}


\bibitem{} {Fan, J. and Li, R.  (2001). Variable selection via nonconcave penalized likelihood and its oracle properties. {\it Journal of the American Statistical Association}, {\bf 96}, 1348--1360.}


\bibitem{} {Frank, J. E. and Friedman, J. H.  (1993). A statistical view of some chemometrics regression tools. {\it Technometrics}, {\bf 35}, 109--148.}


\bibitem{} {Fu, W. J. (1998). Penalized regression: the bridge versus the lasso. {\it Journal of Computational and Graphical Statistics}, {\bf 7}, 397--416.}


\bibitem{} {{Hastie, T. and Tibshirani, R.} (1990). \textit{Generalized Additive Models}.  London: {Chapman $\&$ Hall}.}


\bibitem{} {Hoerl, A. E. and Kennard, R. W.  (1970). Ridge regression: biased estimation for nonorthogonal problems.  \textit{Technometrics}, {\bf 12}, 55--67.}


\bibitem{} {Huang, J., Horowitz, J. L. and Ma, S. (2008). Asymptotic properties of bridge estimators in sparse high-dimensional regression models. {\it Annals of Statistics}, {\bf 36}, 587--613.}


\bibitem{} {Huang, J., Ma, S., Xie, H. and Zhang, C.-H. (2009). A group bridge approach for variable selection. {\it Biometrika}, {\bf 96}, 339--355.}


\bibitem{} {Hurvich, C. M. and Tsai, C.-H. (1989). Regression and time series model selection in small samples. {\it Biometrika}, {\bf 76}, 297--307.}


\bibitem{} {Hurvich, C. M., Simonoff, J. S. and Tsai, C.-H. (1998). Smoothing parameter selection in nonparametric regression using an improved Akaike information criterion. {\it Journal of the Royal Statistical Society Series B}, {\bf 60}, 271--293.}


\bibitem{} {Ishiguro, M., Sakamoto, Y. and Kitagawa, G. (1997). Bootstrapping log likelihood and EIC, an extension of AIC. {\it Annals of the Institute of Statistical Mathematics}, {\bf 49}, 411--434.}


\bibitem{} {Knight, K. and Fu, W. J. (2000). Asymptotics for lasso-type estimators. {\it Annals of Statistics}, {\bf 28}, 1356--1378.}


\bibitem{} {Konishi, S., Ando, T. and Imoto, S. (2004). Bayesian information criteria and smoothing parameter selection in radial basis function networks.  \textit{Biometrika},  \textbf{91}, 27--43.}


\bibitem{} {Liu, Y., Zhang, H., Park, C. and Ahn, J. (2007). Support vector machines with adaptive $L_q$ penalty. {\it Computational Statistics and Data Analysis}, {\bf 51}, 6380--6394.}


\bibitem{} {Lv, J. and Fan, Y. (2009). A unified approach to model selection and sparse recovery using regularized least squares.  \textit{Annals of Statistics},  \textbf{37}, 3498--3528.}


\bibitem{} {McDonald, G. and Schwing, R. (1973). Instabilities of regression estimates relating air pollution to mortality. {\it Technometrics}, {\bf 15}, 463--482.}


\bibitem{} {Park, C. and Yoon, Y. J. (2011). Bridge regression: adaptivity and group selection. {\it Journal of  Statistical Planning and Inference}, {\bf 141}, 3506--3519.}


\bibitem{} {{Schwarz, G.} (1978). {Estimating the dimension of a model.} \textit{Annals of Statistics}, \textbf{6}, {461}--{464}. }


\bibitem{} {Sugiura, N. (1978). Further analysis of the data by Akaike's information criterion and finite corrections. {\it Communications in Statistics - Theory and Methods}, {\bf A7}, 13--26.}


\bibitem{} {{Tibshirani, R.} (1996). {Regression shrinkage and selection via the lasso.} \textit{Journal of the Royal Statistical Society Series B}, \textbf{58}, {267}--{288}. }


\bibitem{} {Tierney, L. and Kadane, J. B. (1986). Accurate approximations for posterior moments and marginal densities. {\it Journal of the American Statistical Association}, {\bf 81}, 82--86.}


\bibitem{} {{Yuan, M. and Lin, Y.} (2006). {Model selection and estimation in regression with grouped variables.} \textit{Journal of the Royal Statistical Society Series B}, \textbf{68}, {49}--{67}. }


\bibitem{} {{Zhang, C.-H.} (2010). {Nearly unbiased variable selection under minimax concave penalty.} \textit{Annals of Statistics}, \textbf{38}, {894}--{942}. }


\bibitem{} {{Zou, H. and Hastie, T.} (2005). {Regularization and variable selection via the elastic net.} \textit{Journal of the Royal Statistical Society Series B}, \textbf{67}, {301}--{320}. }


\bibitem{} {Zou, H. and Li, R. (2008). One-step sparse estimates in nonconcave penalized likelihood models. {\it Annals of Statistics}, {\bf 36}, 1509--1533.}

\end{thebibliography}
\end{document}